\documentclass[%
 aip,
 jcp,
 amsmath,amssymb,
 reprint,%
]{revtex4-1}
\usepackage[utf8]{inputenc}
\usepackage{graphicx}
\usepackage{float}
\usepackage{caption}
\usepackage{subcaption}
\usepackage{xcolor}
\usepackage{dcolumn}
\usepackage{bm}
\usepackage{tabularx}
\usepackage{booktabs}
\usepackage{amsfonts}
\usepackage{mathptmx}
\usepackage{etoolbox}
\usepackage[colorlinks=true,linkcolor=blue,urlcolor=blue,citecolor=blue]{hyperref}
\usepackage{multirow}
\usepackage{todonotes}
\usepackage{microtype}

\newcommand{\fwu}{Center for Advanced Systems Understanding, 02826 G\"orlitz, Germany}
\newcommand{\hzdr}{Helmholtz-Zentrum Dresden-Rossendorf, 01328 Dresden, Germany}

\graphicspath{{figs_main/}}

\makeatletter
\def\@email#1#2{%
 \endgroup
 \patchcmd{\titleblock@produce}
  {\frontmatter@RRAPformat}
  {\frontmatter@RRAPformat{\produce@RRAP{*#1\href{mailto:#2}{#2}}}\frontmatter@RRAPformat}
  {}{}
}%
\makeatother

\begin{document}

\preprint{AIP/123-QED}

\title{Scalable Machine Learning Model for Energy Decomposition Analysis in Aqueous Systems}

\author{Hossein Tahmasbi}
\affiliation{\fwu}
\affiliation{\hzdr}

\author{Michael Beerbaum}%
\affiliation{\fwu}
\affiliation{\hzdr}

\author{Bartosz Brzoza}
\affiliation{\fwu}
\affiliation{\hzdr}

\author{Attila Cangi}
\affiliation{\fwu}
\affiliation{\hzdr}
\email{a.cangi@hzdr.de}

\author{Thomas D. K\"uhne}
\affiliation{\fwu}
\affiliation{\hzdr}
\email{t.kuehne@hzdr.de}


\begin{abstract}
Energy decomposition analysis (EDA) based on absolutely localized molecular orbitals provides detailed insights into intermolecular bonding by decomposing the total molecular binding energy into physically meaningful components. Here, we develop a neural network EDA model capable of predicting the electron delocalization energy component of water molecules, which captures the stabilization arising from charge transfer between occupied absolutely localized molecular orbitals of one molecule and the virtual orbitals of another. Exploiting the locality assumption of the electronic structure, our model enables accurate prediction of electron delocalization energies for molecular systems far beyond the size accessible to conventional density functional theory calculations, while maintaining its accuracy. We demonstrate the applicability of our approach by modeling hydration effects in large molecular complexes, specifically in metal–organic frameworks.
\end{abstract}

\maketitle

\section{Introduction}
The structure and dynamics of hydrogen bonds in liquid water—essential for understanding its unique properties—have been explored in numerous computational studies~\cite{jorgensen1983comparison, poole1992phase, Sprik1996_Ab-initio-, Geissler2001_Autoioniza, laage2006molecular, Khne2009, vega2011simulating, hassanali2013proton, medders2014development, spura2015fly, cisneros2016modeling, ohto2019accessing, elgabarty2020energy, daru2022coupled, folkestad2024understanding}. At ambient conditions, the local structure of water is traditionally described as approximately tetrahedral: each molecule is, on average, bonded to four nearest neighbors, two as donors and two as acceptors~\cite{Stillinger1980, Clark2010, elgabarty2020tumbling}. Thermal fluctuations, however, lead to deviations from this ideal arrangement.

Building on this structural picture, the nature of intermolecular bonding in liquid water has been extensively investigated and consistently validated through a combination of experimental techniques—such as X-ray and neutron diffraction~\cite{soper2000radial, Hura2003, skinner2013benchmark, amann2016x} and vibrational spectroscopy~\cite{bakker2010vibrational, zhang2013vibrational, medders2015infrared, perakis2016vibrational, ojha2018nuclear}—as well as theoretical approaches, including combining first-principles molecular dynamics simulations~\cite{ojha2018hydrogen, elgabarty2019enhancement} with energy-decomposition analysis (EDA) based on absolutely localized molecular orbitals (ALMOs)~\cite{ojha2019time, ojha2021hydrogen}.

In general, EDA is a powerful technique within first-principles electronic structure theory for investigating intermolecular interactions and decomposing the total binding energy into physically meaningful components \cite{hopffgarten2012energy, horn2016probing, pecher2019deriving}. It provides a conceptual bridge between quantum-mechanical calculations and chemical intuition. Unlike conventional EDA methods, the components of ALMO EDA are rigorously defined through constrained variational principles, ensuring a clear and consistent physical interpretation \cite{Khaliullin2007,Khaliullin2008,Khaliullin2013}. This is due to the fact that ALMOs are strictly confined to individual molecules and constructed from their respective atomic orbitals, ensuring both locality and interpretability. 

The ALMO EDA approach for periodic systems~\cite{Khaliullin2013,Khne2013,elgabarty2015covalency} has been implemented in the \texttt{CP2K} code~\cite{Khne2020}, enabling the decomposition of the total interaction energy of water molecules into three distinct components: frozen density interaction, polarization, and electron delocalization.
In addition, ALMOs have been employed in ALMO charge transfer analysis (CTA), a method for quantifying electron density transferred between molecules. Together, ALMO EDA and CTA allow for an accurate determination of charge transfer effects, including intermolecular electron delocalization \cite{elgabarty2015covalency}. This framework provides deeper insight into the nature of hydrogen bonding and other non-covalent interactions in condensed-phase systems  \cite{Khaliullin2013,Khne2013}. Using these techniques, novel aspects of the electronic origins of water’s structure and properties have been uncovered, in particular its asymmetric character~\cite{Khne2014,elgabarty2020tumbling, balos2022time}.

Over the past decade, ALMO EDA has been applied across a wide range of fields, from probing intermolecular interactions in excited states~\cite{Ge2018} to drug design and studies of large molecular systems~\cite{Phipps2016}. More recently, it has been extended to treat open-shell systems~\cite{Horn2013}, metallic systems~\cite{Chen2022}, and to quantify the influence of intermolecular interactions on chemical reactivity in solution~\cite{Mao2021}.

In parallel, machine learning has emerged as a powerful tool in computational physics~\cite{kessler2021artificial,ghasemi2021artificial}, in particular for accelerating property prediction in chemistry and materials science~\cite{Fiedler2022}. The development of interatomic potentials has advanced considerably in methodology, scalability, and software frameworks in recent decades~\cite{Mueller2020_Machine-le}. Foundational approaches such as neural network potentials~\cite{Behler2007_Generalize}, Gaussian Approximation Potentials (GAP)~\cite{Bartok2010_Gaussian-A}, Spectral Neighborhood Analysis Potentials (SNAP)~\cite{thompson_spectral_2015}, Moment Tensor Potentials (MTP)~\cite{Shapeev2016_Moment-Ten}, and the Atomic Cluster Expansion (ACE) framework~\cite{Drautz2019_Atomic-clu} established the basis for subsequent automated learning workflows, including DPMD~\cite{Zhang2018_Deep-Poten}, SchNet~\cite{SchA1/4tt2018_SchNet-a-A}, as well as HIP-NN~\cite{Lubbers2018_Hierarchic}.
These efforts paved the way for the current generation of universal interatomic potential models such as NequIP~\cite{Batzner2022_E3-equivar}, Allegro~\cite{Musaelian2023_Learning-l}, and MACE~\cite{batatia2022mace}.

Beyond interatomic potentials, machine learning has also been applied—though less extensively—to the prediction of electronic structure, particularly in the context of ground-state density functional theory (DFT). This area has recently become an active field of research~\cite{brockherde_bypassing_2017,eickenberg_solid_2018,MiRy2019,grisafi_transferable_2019,ben_mahmoud_learning_2020,ellis_accelerating_2021,FiMoScVo23,madanchi2024simulations,CaFiBrSh25}.

Motivated by these developments, in this work we introduce a neural network EDA model to predict the electron delocalization energy in liquid water. Smooth Overlap of Atomic Positions (SOAP) descriptors are employed to generate atomic fingerprints centered on the oxygen atom of each water molecule \cite{Bartk2013}. The mapping from descriptors to electron delocalization energies is performed using a multi-layer perceptron model. Our model predicts both the first and second strongest donor and acceptor interaction energies and is trained on a large, pre-existing dataset for liquid water~\cite{Khne2014}.

Once trained, the model leverages the locality assumption, analogous to machine learning models for interatomic potentials and electronic structure. This allows us to extend predictions of electron delocalization energies to much larger systems that are otherwise intractable for EDA. We demonstrate the scalability and transferability of our approach by applying it to water molecules confined within the metal–organic frameworks CPO-27-M (M$_{72}$C$_{288}$H$_{72}$O$_{216}$, M = Zn or Cu)~\cite{Klo2024}, which were not included in the training data.

\section{Methods}
\subsection{Datasets}\label{sec:datasets}
Our dataset consists of 5000 snapshots of liquid water at ambient conditions, with each snapshot containing 125 water molecules within a periodic cubic cell of 15.536 Å, totaling 625,000 molecules \cite{Khne2014}. We divided this dataset into 300,000 samples for training, 75,000 samples for validation, and 250,000 for testing.

\subsubsection{Features}
As already indicated, we employed the SOAP descriptor to encode the local atomic environments of each "oxygen-centered" water molecule, transforming them into features that are invariant under translation, rotation, and permutation of like atoms \cite{Bartk2013}. We used the DScribe library~\cite{dscribe,dscribe2} for this task. To optimize the SOAP parameters, we explored the following ranges of values for the cutoff radius (4 to 8 Å), the number of radial basis functions ($nmax$) (8, 10, 12), and the maximum degree of spherical harmonics ($lmax$) (6, 8, 10, 12). Our analysis led to the following final choice of parameters: $nmax=8$, $lmax=6$, and a cutoff radius of 5 \AA, which yields a SOAP feature vector of size $952$ for each molecule. While increasing $nmax$ and $lmax$ generally enhances the descriptor accuracy, it also significantly increases the number of features that can lead to overfitting. 

\subsubsection{Labels}
ALMO EDA separates the total interaction energy of a molecule into three components, the frozen density interaction, polarization, and electron delocalization. In this work, we focus on the electron delocalization energy which represents the sum of two-body donor-acceptor orbital interactions ($\Delta E_{D \rightarrow A}$). These two-body components arise due to delocalization of electrons from the occupied orbitals of donor molecule $D$ to the virtual orbitals of acceptor molecule $A$. The electron delocalization energy per molecule $\Delta E_C$ can be analyzed by considering each water molecule as a donor or as an acceptor,
\begin{equation*}
\Delta E_C = \sum_{N=1}^{Mol} \Delta E_{C \rightarrow N} = \sum_{N=1}^{Mol} \Delta E_{N \rightarrow C}, 
\end{equation*}
where $C$ is the central molecule and $N$ is its neighbors~\cite{Khne2014} and $\Delta E_{N(C) \rightarrow C(N)}$ are two-body delocalization energy components. Here, donor and acceptor refer to a molecule's role in the electron density transfer.

The configurations for the ALMO EDA were obtained from first-principles molecular dynamics simulations using the second-generation Car-Parrinello method~\cite{khne2007, kuhne2018disordered}. 
Reference data generation on ALMO EDA was performed on these 5000 liquid water snapshots. The calculations were previously performed using the \texttt{CP2K} code~\cite{Khne2020}. Further details regarding the methods used for dataset generation are available in the supplementary material of Ref. \cite{Khne2014}.

The delocalization energy per molecule was broken down into its individual interactions, which were then ordered by strength. On average, five strongest donor (acceptor) interactions contribute to the delocalization energy of a molecule. The average of the delocalization energy of a molecule $\langle\Delta E_C\rangle$ is over all central molecules and snapshots (i.e., 625,000 molecular configurations). The electron delocalization is dominated by two strong intermolecular contacts, which together are responsible for about $93 \%$ of the delocalization energy of a single water molecule. The third and the fourth strongest donor (acceptor) interactions contribute about $5 \%$ which indicates the presence of over-coordinated molecules and correspond to back-donation of electrons to (from) the remaining two first-shell neighbors. The remaining $2 \%$ comes from interactions with the second and more distant coordination shells~\cite{Khne2013, Khne2014}. According to these results, on average, each water molecule appears to form two donor and two acceptor bonds. Moreover, a significant asymmetry in the local environment of water molecules is suggested by comparing the strengths of their first and second strongest donor-acceptor interactions~\cite{Khne2013,Khne2014,elgabarty2015covalency,elgabarty2020tumbling}.

\begin{figure}[H]
\centering
\includegraphics[width=1.0\columnwidth]{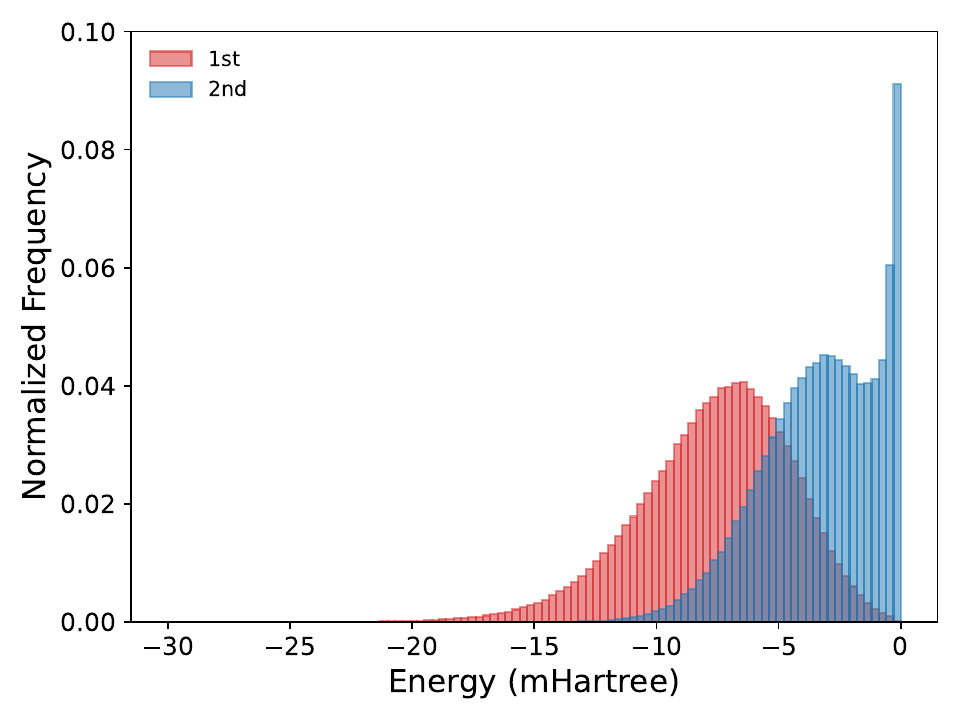}
    \caption{Distribution of the first and second strongest components of the electron delocalization energy used as the dataset of labels. These interactions are responsible for approximately 93\% of a single molecule's total delocalization energy~\cite{Khne2014}.}
\label{fig:output_datasets}
\end{figure}

In this work, we use the first two strongest donor (or acceptor) interaction energies $\Delta E_{C \rightarrow N}$ ($\Delta E_{N \rightarrow C}$) shown in Figure~\ref{fig:output_datasets} as output datasets for the training process of our neural network EDA model. The corresponding energy distributions for acceptor interactions are presented in Figure S1 of the supplementary material~\cite{sup_mat}. 

Also note that prior to training, the raw output data was transformed using logarithmic scaling to reduce skewness and compress the range of values (see Figure S2 of the supplementary material~\cite{sup_mat}). This pre-processing step was done to improve model convergence, stabilize training, and enhance the overall performance of the neural network by ensuring that the data distribution was more suitable for learning.

We also investigated the relationship between our input and output datasets in terms of a similarity analysis with results shown in Figure S3 and Figure S4 of the supplementary material~\cite{sup_mat}.

\subsection{Model training}
We set up a multi-layer perceptron model using a custom Python implementation based on PyTorch \cite{paszke2019pytorch}. To enhance model performance and prevent overfitting, we utilized the Adam optimizer in conjunction with various regularization techniques, including L2 weight decay, dropout, batch normalization, and early stopping. Additionally, we implemented a learning rate scheduler to dynamically adjust the learning rate throughout training, which further improved model performance and convergence. Hyperparameter optimization was conducted using the Optuna library \cite{akiba2019optuna}, enabling systematic tuning of hyperparameters including the learning rate, number of hidden layers, layer widths, batch sizes, and activation function. We employed two distinct samplers \texttt{TPESampler} and \texttt{CmaEsSampler} to guide the optimization process. By using both samplers, we aimed to use their respective strengths and ensure a comprehensive search of the hyperparameter landscape. 

For the learning rate, we searched within a logarithmic range from $10^{-6}$ to $10^{-2}$. The number of hidden layers was allowed to change over integer values within a range from 1 to 40 layers. The width of hidden layers was also optimized, with the number of neurons ranging from 50 to 400. The batch size was optimized over integer values of 16, 32, 64, 128, and 256. The activation functions evaluated included ReLU, Tanh, Sigmoid, and LeakyReLU and the number of epochs was fixed to $50$ as well. A total of 1000 trials were conducted, with each sampler contributing to the overall optimization process, aiming to minimize the validation loss.

Furthermore, for regularization, we investigated the presence or absence of batch normalization and dropout layers with probabilities ranging from 0.0 to 0.9. We also explored the application of weight decay, with values ranging logarithmically from $10^{-5}$ to $10^{-2}$.

After an extensive hyperparameter search, the final network architecture was constructed using SOAP feature vectors of size $952$ as input, two hidden layers each comprising 50 neurons, and output vector of size 2 representing the first and second strongest two-body delocalization energy components. The hyperbolic tangent function was selected as a suitable activation function. Training was carried out using a batch size of $128$, a learning rate of $10^{-4}$, and a weight decay of  $10^{-3}$. It is noteworthy that, in our investigations, the application of batch normalization and dropout layers did not contribute to improving the neural network's performance. 

\section{Results and discussion}
\subsection{Neural network EDA model}
In the initial step, we trained a neural network model designed to predict the first strongest donor interaction, represented as a single output. Figure S5 presents the training and validation loss over epochs, as well as the correlation plot comparing the true and predicted values on the testing dataset comprising 250,000 samples. This model achieved a root mean square error (RMSE) of $0.345$ mHartree on the test dataset, indicating good predictive performance.

\begin{figure}[H]
\centering
\includegraphics[width=1.0\columnwidth]{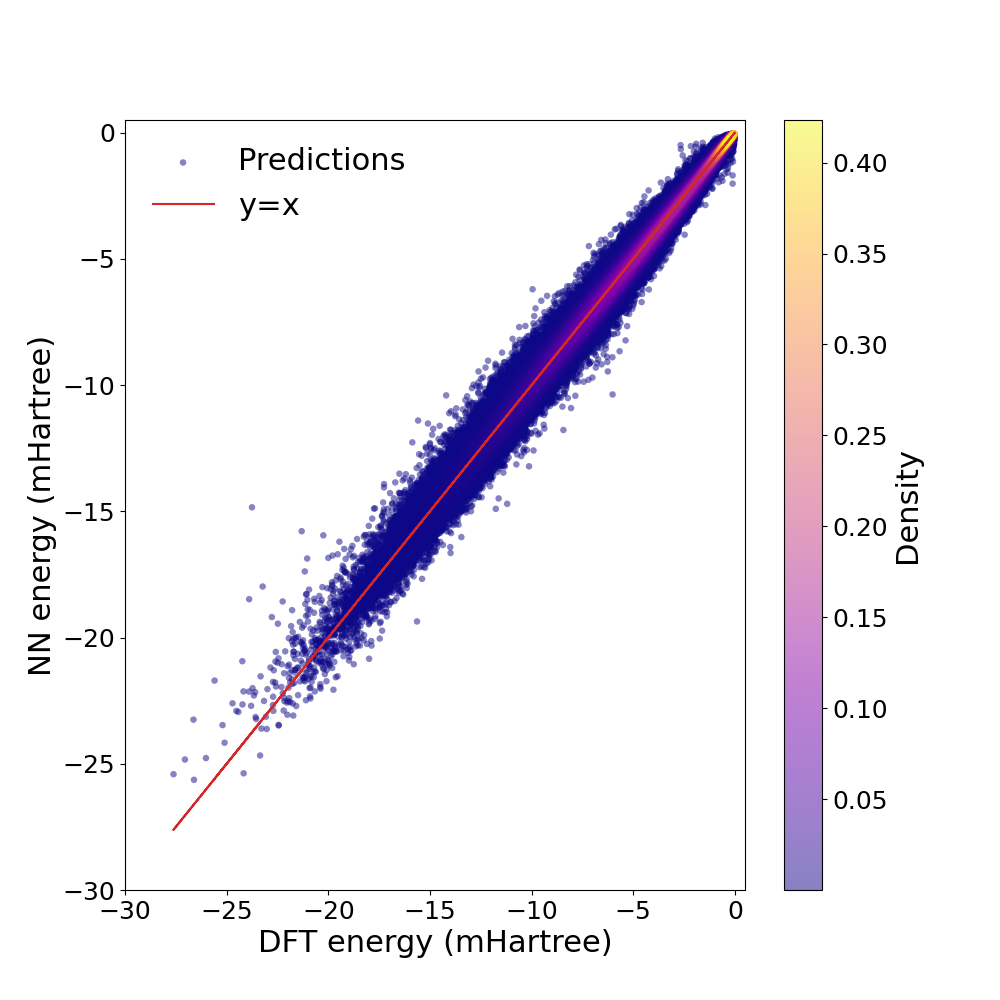}
    \caption{Correlation plot of the true versus predicted energies for the two strongest donor interactions. The color bar indicates the density of data points.}
\label{fig:correlation}
\end{figure}

Building on the initial model, we extended the neural network to predict the two strongest donor interactions simultaneously, using the same architecture. Figure S7 displays the training and validation loss curves over epochs, demonstrating the model’s convergence and generalization capability. Figure~\ref{fig:correlation} shows the correlation plot between the true and predicted values, highlighting the model’s predictive accuracy, with an RMSE of 0.369 mHartree.

\begin{figure}[h!]
\centering
\includegraphics[width=1.0\columnwidth]{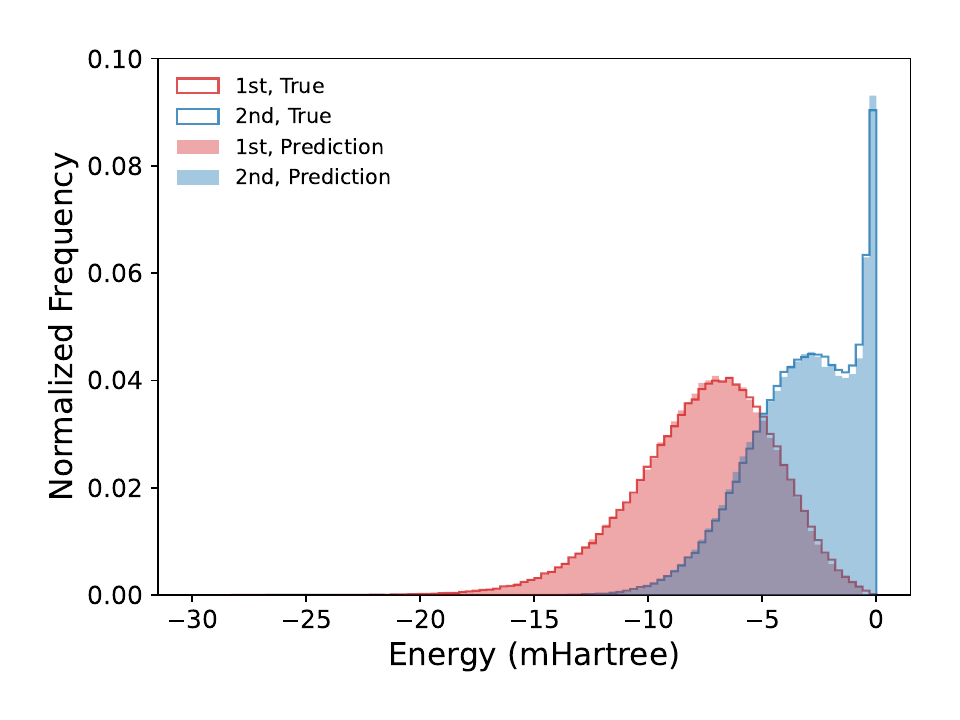}
\caption{Testing the predicted energy distribution of the first two strongest donor interactions. The solid areas show the prediction of the neural network EDA model whereas the solid lines denote the reference data of the test sets.}
\label{fig:dis_true_comp}
\end{figure}

We validated our model by comparing its predictions of the two strongest donor interactions with respect to the test dataset. Figure \ref{fig:dis_true_comp} shows this comparison with a bin size of $100$ to effectively balance resolution and statistical noise, demonstrating the accuracy of our model.

\subsection{Application: Electron delocalization energy of water molecules in metal–organic frameworks}
We demonstrate the capability of our pre-trained model to predict the electron delocalization energy of water molecules in systems that are otherwise computationally expensive or even inaccessible. As a test case, we consider liquid water confined within the metal–organic frameworks CPO-27-M (M$_{72}$C$_{288}$H$_{72}$O$_{216}$, M = Zn or Cu), using datasets from the work of Kloß et al.~\cite{Klo2024}.

Although CPO-27-Zn and CPO-27-Cu are topologically identical, they exhibit markedly different hydration properties. We apply our neural network EDA model to predict the delocalization energies of water molecules within the pores of these frameworks and propose a potential connection between the predicted interaction energies and the experimentally observed differences in hydration behavior. Leveraging our neural network EDA model allows us to avoid the high computational costs typically associated with DFT calculations.

\begin{figure}[H]
\centering
\includegraphics[width=1.0\linewidth]{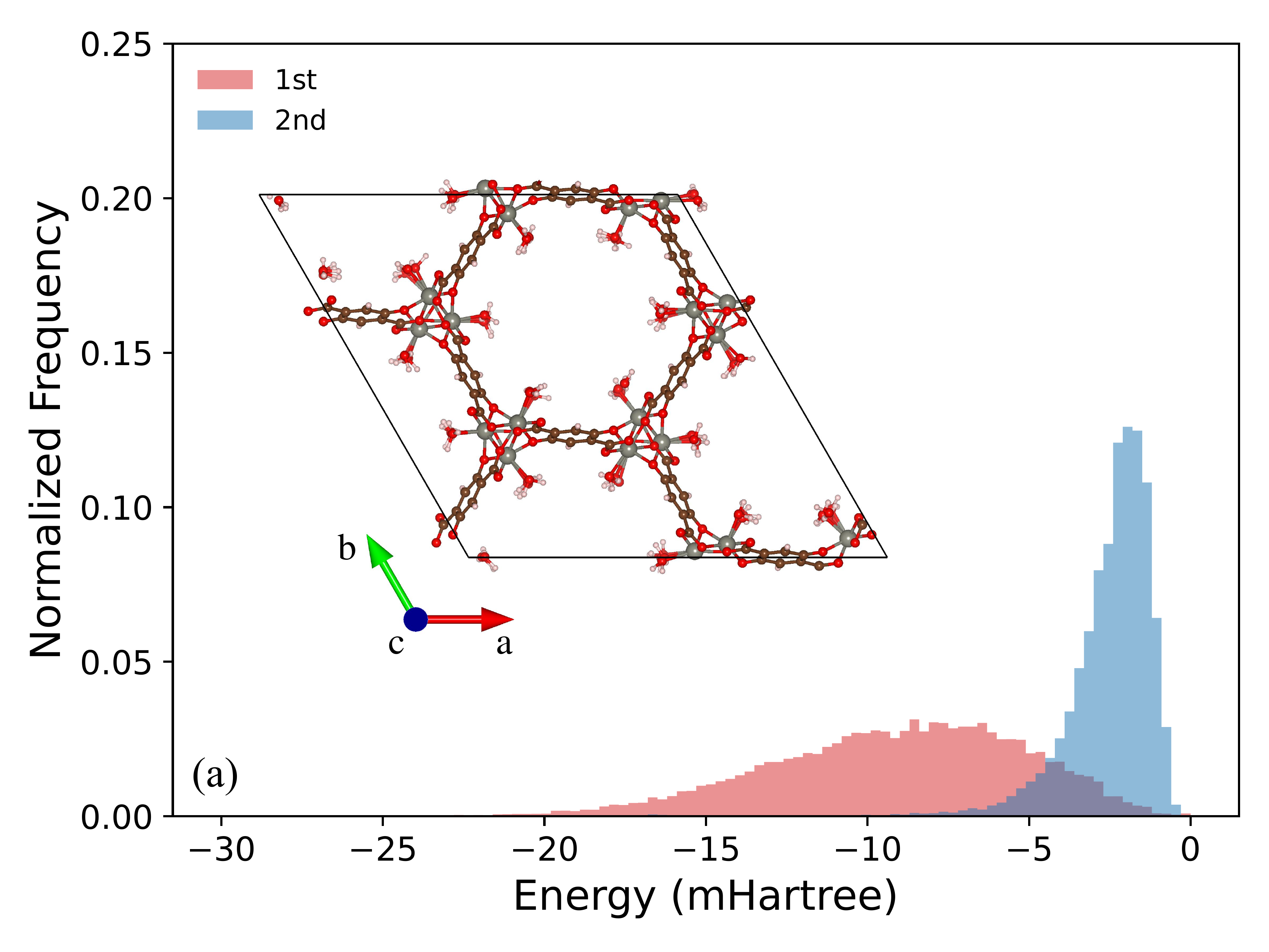}
\includegraphics[width=1.0\linewidth]{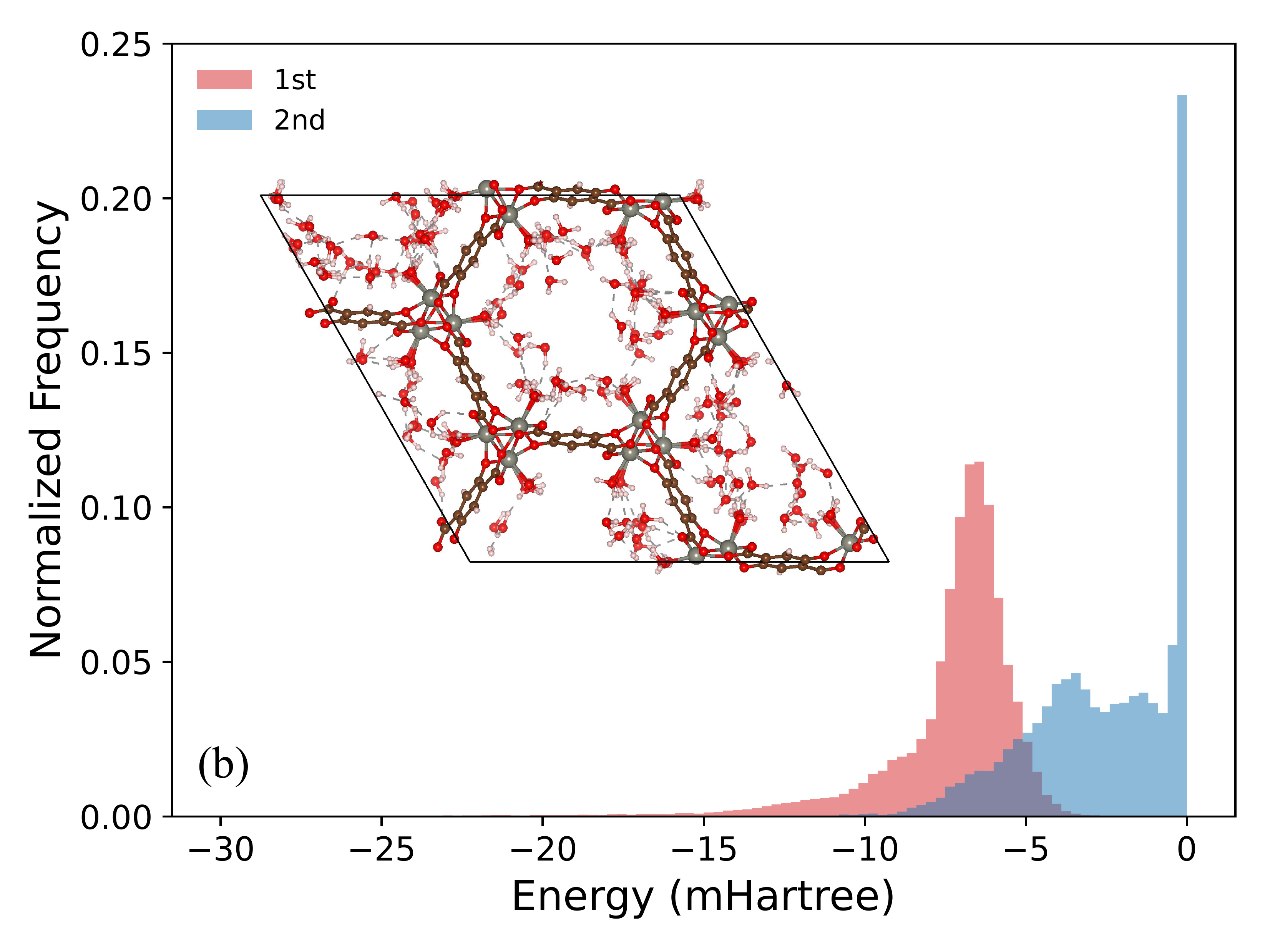}
\includegraphics[width=1.0\linewidth]{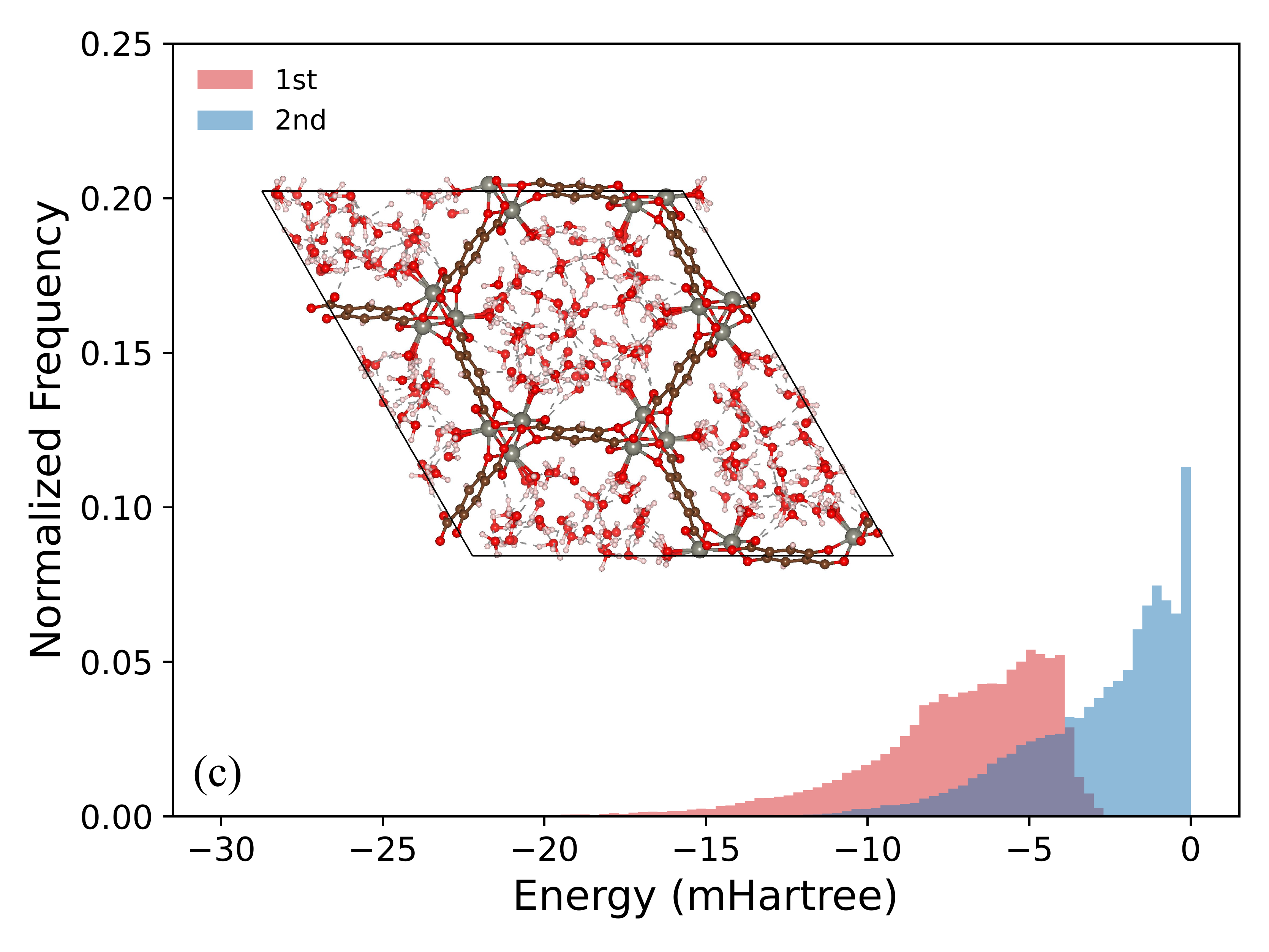}
    \caption{Distribution of the delocalization energy components for water molecules within CPO-27-Zn predicted by the neural network EDA model for: (a) $x=2$, (b) $x=5$, and (c) $x=7$.}
\label{fig:EDA_CPO_Zn}
\end{figure}

\begin{figure}[H]
\centering
\includegraphics[width=1.0\linewidth]{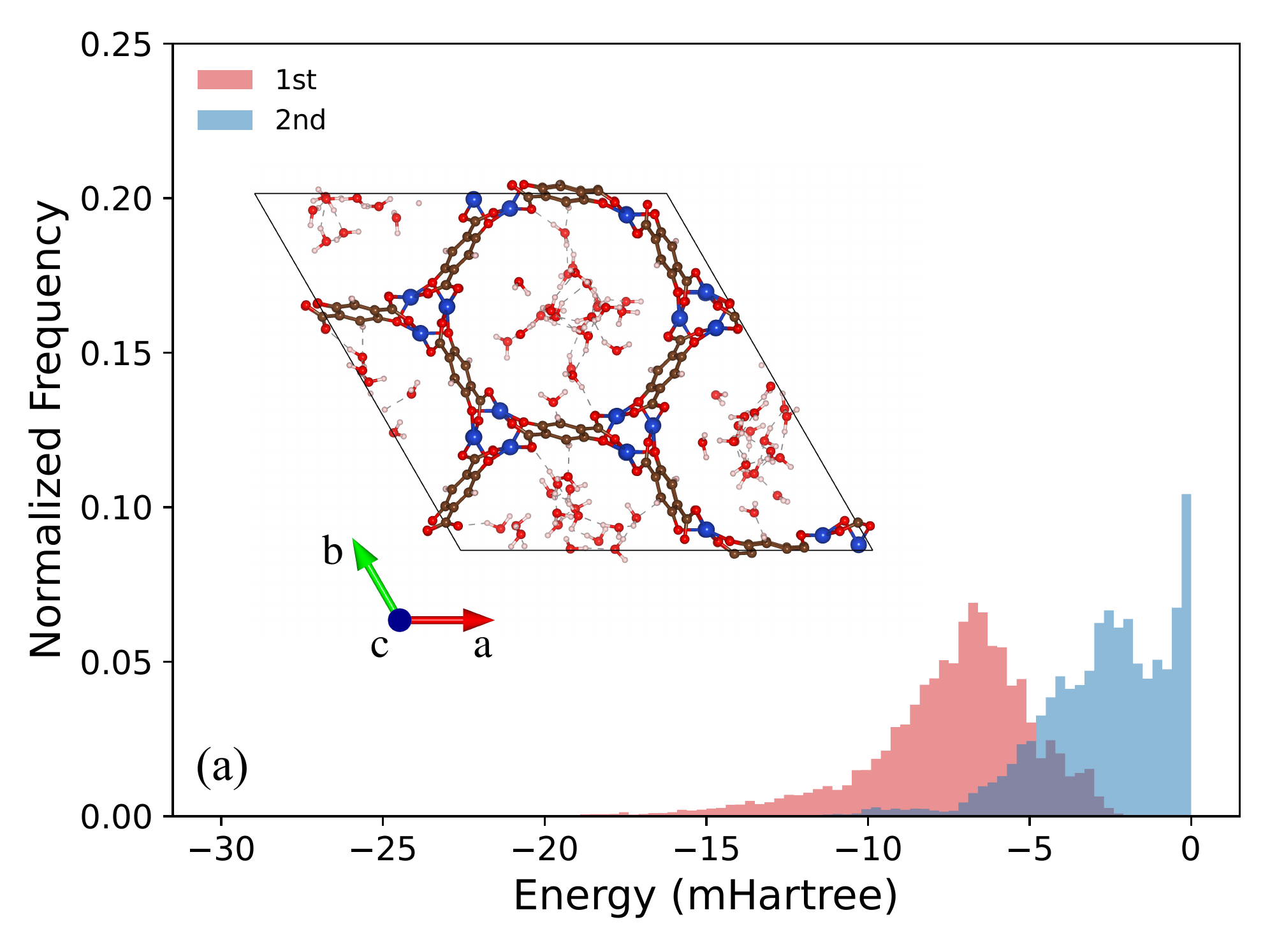}
\includegraphics[width=1.0\linewidth]{Cu2W5.pdf}
\caption{Distribution of the delocalization energy components for water molecules within CPO-27-Cu predicted by the neural network EDA model for: (a) $x=2$ and (b) $x=5$.}
\label{fig:EDA_CPO_Cu}
\end{figure}

The insets of Figure~\ref{fig:EDA_CPO_Zn} illustrate three hydrophilic porous structures CPO-27-Zn. The specific compositions are Zn$_2$(dobdc)(H$_2$O)$_x$ with $x=2,5,7$, containing $72$, $183$ and $258$ water molecules, respectively. Similarly, the insets of Figure~\ref{fig:EDA_CPO_Cu} depict two hydrophobic porous structures CPO-27-Cu, i.e. Cu$_2$(dobdc)(H$_2$O)$_x$ with $x=2,5$, including $72$ and $174$ water molecules, respectively.

For each configuration, we selected one thousand equilibrated snapshots~\cite{Klo2024}. The SOAP features for the water molecules were then generated using the same parameters as for bulk water (see Section~\ref{sec:datasets}). Prior to this step, all other species were removed from the supercells. Finally, the electron delocalization energies of all water molecules were computed using our neural network EDA model. The resulting energy distributions for each system are shown in Figure~\ref{fig:EDA_CPO_Zn} and Figure~\ref{fig:EDA_CPO_Cu}, where each plot represents the delocalization energies of all water molecules across the complete set of 1000 snapshots per supercell.

As shown in Figure~\ref{fig:EDA_CPO_Zn}, the strong hydrophilic behavior of Zn$_2$(dobdc) binds water molecules to the free coordination sites. With increasing water content, a robust hydrogen-bonded network forms, extending from the pore walls toward the pore centers. The charge on the zinc atoms is redistributed through this network, resulting in a positively charged surface at the pore walls. This highlights how the hydrophilic nature of Zn$_2$(dobdc) promotes extended water structuring within the pores.

In contrast, the hydrophobic character of Cu$_2$(dobdc) leads to the accumulation of water molecules in the pore centers (Figure~\ref{fig:EDA_CPO_Cu}). Unlike the zinc-based framework, no extended hydrogen-bonded network emerges from the pore walls. Instead, interactions with the framework occur primarily through hydrogen bonding at the interface, and no significant charge transfer is observed~\cite{Klo2024}. This stark difference underscores how subtle changes in the framework composition (Zn versus Cu) fundamentally alter the hydration behavior and electron delocalization profiles of confined water.

The confinement within the pores, the hydrophobic/hydrophilic character of the framework, and the number of confined water molecules strongly influence the resulting energy distributions. In the case of hydrophilic pores, as in Zn$_2$(dobdc) (Figure~\ref{fig:EDA_CPO_Zn}), several characteristic shifts are observed. Isolated water molecules, as in Zn$_2$(dobdc)(H$_2$O)$_2$, show distributions shifted toward lower energies in both the strongest and second-strongest donor interactions. Notably, the signal at $0{-}0.3$ mHartree for the second-strongest donor interaction is absent in this isolated case. In contrast, this signal becomes dominant once extended water networks begin to form, as in Zn$_2$(dobdc)(H$_2$O)$_5$. With increasing hydration, these networks generate pronounced asymmetric shifts of the distributions toward lower energies, and the asymmetry is further enhanced at higher water loadings, such as Zn$_2$(dobdc)(H$_2$O)$_7$.

By contrast, the hydrophobic character of Cu$_2$(dobdc) drives the formation of a water column in the pore center (Figure~\ref{fig:EDA_CPO_Cu}). As the number of confined molecules increases, the density within the column rises, leading to markedly different energy distributions compared to the zinc-based framework. Unlike Zn$_2$(dobdc), Cu$_2$(dobdc) does not exhibit strong asymmetric shifts. The signal at $0-0.3$ mHartree is comparatively weaker than in bulk water and is strongly enhanced in Zn$_2$(dobdc). In Cu$_2$(dobdc)(H$_2$O)$_5$, the addition of water molecules yields a slightly asymmetric peak, where the signal at 0 mHartree is reduced relative to lower pore fillings or hydrophilic environments. As hydration increases further, confinement effects alter bond lengths and angles, which in turn broaden the distributions.

In bulk water, Khaliullin and Kühne attributed the signal around 0 mHartree for the second-strongest donor interaction to configurations in which donation to the second acceptor is weaker than back-donation~\cite{Khaliullin2013}. This back-donation signal becomes strongly enhanced when water is confined in the metal–organic frameworks Zn$_2$(dobdc)(H$_2$O)$_5$ and Zn$_2$(dobdc)(H$_2$O)$_7$, where water molecules bound at the pore surface carry a high charge~\cite{Klo2024}. Redistribution of this charge across the hydrogen-bonded network further amplifies the signal. In addition, water molecules in Zn$_2$(dobdc) exhibit pronounced position- and orientation-dependent behavior, which results in narrower distributions compared to bulk water. These variations effectively create distinct “species” of confined water, each with a characteristic energy profile. As the number of confined molecules increases, the diversity of these species grows, leading to less optimal geometries and consequently greater overlap and broadening of the distributions.

\section{Conclusions}
In this work, we developed a supervised machine learning model based on multi-layer perceptrons for an EDA of liquid water. Our neural network EDA model is designed to predict the first two strongest components of the electron delocalization energy for a given atomic environment around each water molecule. By leveraging SOAP descriptors as features, the model maps the local atomic neighborhood of oxygen-centered water molecules to their corresponding electron delocalization energies.

Similar to machine learning interatomic potentials, our approach exploits the principle of locality. This assumption enables the training of a pre-trained model on systems where reference EDA data is available and its subsequent application to much larger systems that would otherwise be computationally inaccessible using conventional EDA methods.

We demonstrated the capabilities of our neural network EDA model by predicting and interpreting the hydration behavior of water confined within metal–organic frameworks. In doing so, we showed that our model not only reproduces key features of electron delocalization in bulk water but also provides insight into the distinct hydration properties of hydrophilic and hydrophobic pore environments.

While the present model has been trained on liquid water, the methodology is broadly transferable to aqueous systems. By combining locality-based descriptors with supervised neural network models, this framework sets the stage for constructing a universal machine learning EDA model applicable to a wide range of molecular systems and condensed-phase environments.


\section*{Acknowledgments}
This work was supported by the Center for Advanced Systems Understanding (CASUS), which is financed by Germany’s Federal Ministry of Research, Technology and Space (BMFTR) and by the Saxon State government out of the State budget approved by the Saxon State Parliament. 
Computations were performed on a Bull Cluster at the Center for Information Services and High-Performance Computing (ZIH) at Technische Universit\"at Dresden and on the cluster Hemera of the Helmholtz-Zentrum Dresden-Rossendorf (HZDR).

\section*{Author Declarations}

\subsection*{Conflict of interest}
The authors have no conflicts to disclose.

\section*{Data Availability Statement} 
The data that support the findings of this study are available upon request from the authors.

\section*{References}
\nocite{*}
\bibliography{main}

\end{document}